\documentclass[aps,showpacs,groupedaddress,nofootinbib,letterpaper,preprint]{revtex4}

\usepackage{amsmath}
\usepackage{amsfonts}
\usepackage{amssymb}

\usepackage{hyperref}

\newcommand{\avg}[1]{\left\langle #1 \right\rangle}
\newcommand{\bra}[1]{\left\langle #1 \right|}
\newcommand{\ket}[1]{\left| #1 \right\rangle}

\newcommand{\cbar}{\overline{c}}

\begin{document}
\title{Do gauge fields really contribute negatively to black hole entropy?}

\author{William Donnelly}
\affiliation{
Center for Fundamental Physics, Department of Physics, \\
University of Maryland, College Park, Maryland 20742-4111, USA
}
\email{wdonnell@umd.edu}
\author{Aron C. Wall}
\affiliation{
Department of Physics, \\
University of California, Santa Barbara \\
Santa Barbara, CA 93106, USA
}
\email{aroncwall@gmail.com}

\begin{abstract}
Quantum fluctuations of matter fields contribute to the thermal entropy of black holes.
For free minimally-coupled scalar and spinor fields, this contribution is precisely the entanglement entropy.
For gauge fields, Kabat found an extra negative divergent ``contact term'' with no known statistical interpretation. 
We compare this contact term to a similar term that arises for nonminimally-coupled scalar fields.
Although both divergences may be interpreted as terms in the Wald entropy, we point out that the contact term for gauge fields comes from a gauge-dependent ambiguity in Wald's formula.
Revisiting Kabat's derivation of the contact term, we show that it is sensitive to the treatment of infrared modes.
To explore these infrared issues, we consider two-dimensional compact manifolds, such as Euclidean de Sitter space, and show that the contact term arises from an incorrect treatment of zero modes.
In a manifestly gauge-invariant reduced phase space quantization, the gauge field contribution to the entropy is positive, finite, and equal to the entanglement entropy.
\end{abstract}

\pacs{
04.70.Dy 	
11.15.-q 	
04.60.Kz  
11.10.Gh 	
}

\maketitle

\tableofcontents

\section{Introduction}

The entropy of a horizon is given to leading order by the Bekenstein-Hawking formula $S_\mathrm{BH} = A/4G$ ($c = \hbar = 1$).
However, the quantum fields near the horizon are in a highly entangled state, and their entropy  $S_\mathrm{ent} = -\mathrm{tr}(\rho\,\ln \rho)$ should contribute to the Bekenstein-Hawking entropy.
It is well known \cite{Bombelli1986,Srednicki1993} that the leading-order divergence of the entanglement entropy $S_\mathrm{ent}$ scales as the area $A$ of the horizon.
However, $S_\mathrm{ent}$ also depends on the number of each kind of particle species and their interactions, whereas the Bekenstein-Hawking entropy depends only on Newton's constant $G$.   
This ``species problem'' would be elegantly solved if the number of species affected the renormalization of $1/G$, so that $S_\mathrm{BH}$ implicitly depends on the field content \cite{Jacobson1994,Susskind1994}.

This solution can only work if the fields' entanglement entropy divergence matches their renormalization of $1/G$.  For minimally coupled scalars and spinors the matching is exact, at least at one loop \cite{Fursaev1994,Kabat1995,deAlwis1995,Larsen1995,Demers1995,Winstanley2000}.\footnote{For scalars in odd-dimensional spacetimes one calculation \cite{Kim1996} found a discrepency.  In fact they did not even recover the standard leading order area divergence.  We suspect that this may be due to the fact that the entanglement entropy, unlike most physical observables, is sensitive to arbitrarily high energy states and therefore receives contributions from near the cutoff.  This means that a) different regulator schemes need not commute past one another, and b) because the Pauli-Villars regulator permits negative normed states, there can be large, negative, unphysical contributions to $S$.}
Since entropy is an intrinsically positive quantity, as one flows to the infrared, this requires a positive contribution to $1/G$, resulting in screening of Newton's constant (i.e. gravity falls off with distance faster than would be expected in classical physics).

However, there appears to be a discrepancy for gauge fields.  
This discrepancy was first identified by Kabat \cite{Kabat1995}, who found an extra ``contact term'' divergence in $1/G$ for spin-1 Maxwell fields, which does not correspond to the divergence in the entanglement entropy.  
Similar issues arise for linearized gravity \cite{Fursaev1996}, but here we will focus on the simpler case of Maxwell theory.

The entropy can be calculated by the conical method \cite{Susskind1993,Callan1994} (see Ref.~\cite{Solodukhin2011} for a review). 
Let $Z(\beta)$ be the Euclidean partition function on a spacetime with a conical singularity at the horizon with conical angle $\beta$.
The conical entropy is given by 
\begin{equation} \label{SfromZ}
S_\mathrm{cone} = \left( 1 - \beta \frac{\partial}{\partial \beta} \right) \ln Z(\beta) \Big|_{\beta = 2\pi}.
\end{equation}
The one-loop partition function can be calculated using a heat kernel regulator, and is given by $\ln Z = - \int \sqrt{g} L_\mathrm{eff}$ where
\begin{equation} \label{Leff}
L_\mathrm{eff} = -\frac{1}{2} \int^{+\infty}_{\epsilon^2} 
ds \frac{e^{-sm^2}}{(4\pi s)^{D/2}} \left( \frac{c_0}{s} + c_1 R + \mathcal{O}(s) \right),
\end{equation}
with $s$ a Schwinger proper time coordinate, $\epsilon$ an ultraviolet cutoff, and $m$ the mass.
The coefficient $c_1$ associated with renormalization of $1/G$ also determines the coefficient of the leading-order entropy divergence 
using Eq.~\eqref{SfromZ}:
\begin{equation} \label{Sc1}
S = 2 \pi A c_1 \int_{\epsilon^2}^\infty ds \frac{e^{-m^2s}}{(4 \pi s)^{D/2}}.
\end{equation}

For a minimally coupled scalar field, $c_1 = 1/6$.  For a Maxwell field, the entropy divergence coefficient was found to be \cite{Kabat1995}
\begin{equation} \label{c1}
c_1 = \frac{D - 2}{6} - 1.
\end{equation}
In addition to the $D - 2$ bosonic degrees of freedom which contribute just like $D - 2$ scalars, there is an additional contact term that makes the entropy negative for $D < 8$. 
Negative values of $c_1$ correspond to antiscreening of Newton's constant by Maxwell fields.

This contact term is surprising for a number of reasons.  First, the term seemingly appears as a UV divergence even in $D = 2$, in which Maxwell fields have no local degrees of freedom.\footnote{More generally, for $D \ne 4$ it is not invariant under electric-magnetic duality which relates a massless $p$-form field to a massless $(D - p - 2)$-form field (up to issues involving zero modes of the fields).  For example, in $D = 3$ the on-shell dynamics of the Maxwell field are exactly the same as a massless scalar field, and yet their contributions to the entropy divergence are not even of the same sign.}
In what follows we will work primarily in $D = 2$.  
The reason is that in Kabat's original calculation the polarizations in which the vector is transverse to the horizon contribute exactly like $D-2$ minimally coupled scalar fields, while the contact term comes from the theory reduced to the remaining two dimensions normal to the horizon.
Thus the effect in higher dimensions is closely related to the effect in $D = 2$ dimensions.  

Secondly, because the entanglement entropy $S_\mathrm{ent}$ cannot be negative, one cannot explain the contact term by means of the entanglement entropy divergence.  
A similar discrepancy occurs for a nonminimally coupled scalar field, for which $c_1 = 1/6 - \xi$, but the leading order entanglement entropy divergence is independent of $\xi$.
(This can be seen most easily in flat space where there is no curvature to couple to.)

However, this nonminimal scalar discrepancy can be explained \cite{Kabat1995b,Frolov1997} if we add to the Bekenstein-Hawking entropy the Wald entropy \cite{Wald1993,Visser1993,Jacobson1993,Iyer1994} associated with the nonminimal coupling of the scalar \cite{Jacobson1993}
\begin{equation}
S_\mathrm{Wald}^{(\phi)} = - 2\pi \xi \int_\Sigma d^{D-2}x\,\sqrt{h}\, \phi^2,
\end{equation}
where $\Sigma$ is the bifurcation surface of the horizon, and $h$ is the determinant of the induced metric on $\Sigma$.
This term in the entropy is a consequence of the scalar coupling directly to the singular curvature at the tip of the cone \cite{Solodukhin1995}.
In the quantum theory, $\phi^2$ is divergent and therefore also contributes to the renormalization of $1/G$.  In section \ref{nonmin} we will show how this exactly accounts for the $\xi$-dependent term.

It has been suggested \cite{Larsen1995} that the Kabat contact term can be attributed to an effective nonminimal coupling of the vector field to gravity.
Following the same method as for the nonminimally coupled scalar, the divergent term in the Wald entropy would take the form 
\begin{equation}
S_\mathrm{Wald}^{(A)} = -\pi \int_\Sigma \sqrt{h} \,A_a A_b g^{ab}_\perp
\end{equation} where $g^{ab}_\perp$ is the inverse metric perpendicular to the horizon \cite{privateKabat}.
In section \ref{aint} we will verify that this term can indeed explain the extra renormalization of $1/G$ in Feynman gauge.  However, this term is not manifestly gauge-invariant, nor is it invariant under BRST symmetry.
Furthermore, this term actually corresponds to a Jacobson-Kang-Myers \cite{Jacobson1993} ambiguity in the derivation of the Wald entropy as a Noether charge.  These JKM ambiguities vanish for classical fields at the Killing horizon, but may have nonvanishing quantum expectation values.

In section \ref{kabat}, we review Kabat's derivation of the contact term \cite{Kabat1995}.
In this derivation, the gauge-fixed Maxwell action is integrated by parts in order to put it in the form field-operator-field.  Because of the JKM ambiguity, the Wald entropy can depend on an integration by parts. So one might wonder whether the contact term comes from an improper treatment of the boundaries at the conical singularity and/or infinity.  To eliminate these boundaries, in section \ref{zeromodes} we will consider the analogue of the contact term for two-dimensional smooth compact spacetimes.
The partition function of a cone should be obtainable as a limit of the partition function of smooth, compact geometries.
Indeed we will see that the contact term persists even in the compact case.
However, in this case the contact term in the entropy comes entirely from the zero mode sector.
We will argue that these zero modes have not been properly treated, so that in this case the contact term should be viewed as unphysical.

In section \ref{physical2D}, we calculate the physical partition function on a 2D compact manifold in the reduced phase space, without the use of gauge-fixing or ghosts, taking into account all non-perturbative effects.
We find that the entropy is finite and equal to the entanglement entropy.
As expected on physical grounds, there is no renormalization of $1/G$ in two dimensions.

Thus we conclude that---at least in two dimensions---Maxwell fields do not antiscreen Newton's constant.  
The Kabat contact term is not present in the horizon entropy, and appears to be purely a gauge artifact.
We discuss the implications of this result in section~\ref{dis}.

\section{Contact term as renormalization of Wald entropy?}\label{renWald}

The entropy of a bifurcate Killing horizon can be calculated in a $D$-dimensional diffeomorphism-invariant classical theory by the Wald Noether charge method \cite{Wald1993, Iyer1994}. If the Lagrangian $L$ does not depend on derivatives of the Riemann tensor, and all derivatives of the matter fields are symmetrized, the Wald entropy is given by differentiating $L$ with respect to the Riemann tensor \cite{Visser1993,Jacobson1993}:
\begin{equation} \label{wald}
S_\mathrm{Wald} = -2\pi \int_\Sigma d^{D-2}x\,\sqrt{h}\,\frac{\partial L}{\partial{R_{abcd}}} \epsilon_{ab} \epsilon_{cd},
\end{equation}
where $h$ is the pullback of the metric to the $D-2$ dimensional bifurcation surface $\Sigma$ and $\epsilon_{ab}$ is the binormal to the slice.  This formula was proven to be equivalent to the ``Noether charge'' on stationary Killing horizons.  However, on nonstationary backgrounds, Eq. (\ref{wald}) is ambiguous, since it can be affected by integrating the action by parts, or by performing field redefinitions that involve the metric.

The Wald entropy is classical, and we are interested in the full entropy as defined by the conical entropy formula \eqref{SfromZ}.
For a classical theory, the conical entropy is equivalent to the Wald
entropy \cite{Nelson1994}, while for minimally coupled scalar and spinor fields it equals the entanglement entropy \cite{Kabat1994}.
It is thus natural to conjecture, in accordance with the arguments of Refs.~\cite{Nelson1994,Frolov1997}, that for a general quantum field theory the conical entropy is given by the sum
\begin{equation} \label{conj}
S_\mathrm{cone} = \avg{S_\mathrm{Wald}} + S_\mathrm{ent},
\end{equation}
where the Wald term comes from the coupling of fields to the curvature at the tip of the cone, while the entanglement term comes from the angle deficit away from the tip.  For general relativity with minimally coupled matter, the right-hand side of Eq.~\eqref{conj} is the generalized entropy, which is conjectured to obey the generalized second law \cite{Wall2009}.\footnote{The generalized second law has been proven in various regimes \cite{Wall2009, Wall2010, Wall2012} for fields minimally coupled to general relativity, where $S_\mathrm{Wald} = A/4G$.  For higher curvature gravity and nonminimal couplings, it is not even known whether the theory obeys a classical second law, except for special cases such as $f(R)$ gravity \cite{Jacobson1995}, nonminimally coupled scalars \cite{Ford2000}, and first order perturbations to Lovelock horizons \cite{Kolekar2012}.  However, it is also known that the Wald entropy can decrease in classical mergers of Lovelock black holes \cite{Jacobson1993,Liko2007,Sarkar2010}}

Since $S_\mathrm{cone}$ is defined in terms of the renormalized effective action $-\ln Z$, it must be independent of the renormalization scale.
Therefore an important consistency check of Eq.~\ref{conj} is whether the generalized entropy is also invariant under the renormalization group flow.
This is nontrivial, since the terms $S_\mathrm{Wald}$ and $S_\mathrm{ent}$ depend explicitly on the renormalization scale: the latter because of the ultraviolet divergence of $-\mathrm{tr}(\rho\,\ln\,\rho)$, and the former because of the RG flow of the coupling constants such as $1/G$, and (in some cases) divergent products of fields such as $\phi^2$.  In order for Eq.~\eqref{conj} to hold, the renormalization of the entanglement entropy must match the renormalization of the Wald entropy, when both are regulated in the same way.  We will now check this using the heat kernel regulator for the nonminimally coupled scalar, and for Maxwell theory.

\subsection{Nonminimally coupled scalar field}\label{nonmin}

An illustrative example of a field theory with a consistent contact term is the nonminimally coupled scalar field \cite{Solodukhin1995}.  Its action is
\begin{equation} \label{Inonmin}
I = \int d^Dx\,\sqrt{g} \frac{1}{2} \phi (-\nabla^2 + \xi R) \phi.
\end{equation}
Its contribution to the generalized entropy \eqref{conj} is given by 
\begin{equation} \label{sgennonmin}
S_\mathrm{gen} = - 2\pi \xi \int d^{D-2}x\,\sqrt{h}\, \avg{\phi^2} + S_\mathrm{ent}.
\end{equation}

The coefficient of the entropy divergence is given by $c_1 = 1/6 - \xi$.
The value of $1/6$ comes from the usual entanglement entropy divergence in $S_\mathrm{ent}$, which in flat spacetime is independent of $\xi$.  However, there is also a contact term coming from divergences in $\langle \phi^2 \rangle$.  Divergences of this term are not associated with the entanglement entropy.  Instead they correspond to particle loops that interact with the curvature at the conic singularity.\footnote{This is not necessarily inconsistent with the hypothesis \cite{Jacobson1994} that the horizon entropy ultimately comes entirely from entanglement entropy.  It could be that the nonminimal coupling term is induced by entanglement at an even higher energy scale, as in the $O(N)$ model considered in Ref.~\cite{Kabat1995b}.}  
Although $\phi^2$ is an intrinsically positive quantity, the coupling $\xi$ can take either sign.  
Positive $\xi$ corresponds to antiscreening of Newton's constant.  
In fact there is exact numerical consistency between the divergences in  Eq.~\eqref{sgennonmin} and the renormalization of $1/G$.

The partition function of the theory \eqref{Inonmin} is a functional determinant,
\begin{equation}
\ln Z = -\frac{1}{2} \ln \det \Delta_0^\xi,
\end{equation}
where $\Delta_0^\xi = (-\nabla^2 + \xi R)$.
In the heat kernel regularization, the functional determinant is expressed in terms of the trace of the heat kernel,
\begin{equation}
K_S^\xi(s) = \mathrm{tr} \, e^{- s \Delta_0^\xi}.
\end{equation}
The partition function is then given by
\begin{equation}
\ln Z = \frac{1}{2} \int_{\epsilon^2}^\infty ds \, \frac{e^{-m^2 s}}{s} K_S^\xi(s)
\end{equation}
where $m$ is the mass of the field, and $\epsilon$ is an ultraviolet cutoff with dimensions of length.

The heat kernel can be expressed as a Schwinger path integral over paths $x^a(s)$ through the Euclidean spacetime with $s$ as the ``time'' parameter:
\begin{equation}
K^\xi_\text{S}(s,x,y) = \int_{x(0) = x}^{x(s) = y} \mathcal{D}x e^{\textstyle -\int_0^s ds' \tfrac{1}{4} \dot x^a(s') \dot x_a(s') + \xi R}.
\end{equation}
The heat kernel and its trace are related by
\begin{equation}
K_S^\xi(s) = \int d^Dx \sqrt{g} \, K_S^\xi(s,x,x).
\end{equation}
We are interested in the heat kernel for first-order variations of $\beta$ away from $2 \pi$. 
The conical deficit introduces a singular curvature at the tip, given by \cite{Fursaev1995}
\begin{equation}
R_\mathrm{tip}(x) = 2 (2 \pi - \beta) \delta_\Sigma (x).
\end{equation}
The Schwinger path integral depends on $\beta$ through both the change of angular periodicity, and the introduction of curvature at the tip.
To first order in $\beta - 2\pi$, these two contributions are independent and we can write the trace of the heat kernel as a sum of paths that do not interact with the singularity, and those that do \cite{Solodukhin1995}:
\begin{equation}
K^{\xi}_S(s) = K^{\xi}_S(s)|_{\partial_n \phi = 0} + K_\mathrm{tip}(s).
\end{equation}
The first term is the heat kernel with Neumann boundary conditions $\partial_n \phi = 0$ at the tip of the cone, whose contribution to $S_\mathrm{cone}$ is the entanglement entropy $S_\mathrm{ent}$ \cite{Solodukhin1996}.
The second term is
\begin{align}
K_\mathrm{tip}(s) 
&= \int d^D x \sqrt{g} \int_{x(0) = x}^{x(s) = x} \mathcal{D}x \int_0^s d s' \left( -\xi R_\mathrm{tip}(x(s')) \right) e^{ \textstyle -\int_0^s ds'' \tfrac{1}{4} \dot x^a(s'') \dot x_a(s'') + \xi R} \\
&= -2 \xi (2\pi - \beta) \int d^Dx \sqrt{g} \int_\Sigma d^{D-2}y \sqrt{h} \int_0^s d s' K^\xi_S(s', x, y) K^\xi_S(s - s', y, x) \\
&= -2 \xi (2\pi - \beta) \int_\Sigma d^{D-2}y \sqrt{h} \, s \, K^\xi_\text{S}(s, y, y).
\end{align}
where we have used the heat kernel identity
\begin{equation}
K^\xi_S(s,x,x) = K^\xi_S(s',x,y) K^\xi_S(s - s',y,x).
\end{equation}
The contribution to the effective action is
\begin{align}
\ln Z_\text{tip} 
&= \frac{1}{2} \int_{\epsilon^2}^\infty ds  \, \frac{e^{-m^2 s}}{s} K_\text{tip}(s) \\
&= -\xi (2 \pi - \beta) \int_\Sigma d^{D-2}y \sqrt{h} \int_{\epsilon^2}^\infty ds \, e^{-m^2 s} K^\xi_\text{S}(s,y,y).
\end{align}
We can identify in this last expression the expectation value of $\phi^2$ in heat kernel regularization:
\begin{equation}
\avg{\phi^2(y)} = \int_{\epsilon^2}^\infty ds  \, e^{-m^2 s} K_S^\xi(s,y,y).
\end{equation}
We therefore have
\begin{equation}
\ln Z_\text{tip} = -\xi (2 \pi - \beta) \int_\Sigma d^{D-2}y \sqrt{h} \avg{\phi(y)^2},
\end{equation}
and the contribution to the conical entropy is
\begin{equation} \label{Scurv}
S_\text{tip} = (1 - \beta \partial_\beta) \ln Z_\text{tip} \big|_{\beta = 2 \pi} = -2 \pi \xi \int_\Sigma d^{D-2}y \sqrt{h} \avg{\phi(y)^2}.
\end{equation}
This is precisely the same as the expectation value of the scalar contribution to the Wald entropy \eqref{wald}.
So we see that the conjecture \eqref{conj} holds in the case of the nonminimally coupled scalar field.

\subsection{Maxwell field}\label{aint}

It is tempting to interpret the Maxwell contact term in the same way, as a contribution coming from the Wald entropy, just as in the case of the nonminimally coupled scalar field.

In Ref.~\cite{Kabat1995}, the thermal entropy of Maxwell fields in Rindler space was obtained from the partition function $Z$ on the cone.  The Euclidean action for the Maxwell field includes (fermionic-scalar) ghosts and a gauge fixing term.  In Feynman gauge:
\begin{equation}\label{maxwell}
I = \int d^Dx\,\sqrt{g}\left[\frac{1}{4} F_{ab}F^{ab} + \frac{1}{2}(\nabla_a A^a)^2
- \bar{c}\,\nabla^2\,c \,\right].
\end{equation}
To express the one-loop effective action as a determinant, we integrate by parts:
\begin{equation} \label{action}
I = \int d^Dx\,\sqrt{g}\left[ -\frac{1}{2}A^a (g_{ab} \nabla^2 - R_{ab}) A^b
- \bar{c}\,\nabla^2\,c \right].
\end{equation}
The gauge-fixed vector field $A_a$ has $D$ degrees of freedom, while the two Faddeev-Popov ghosts $c$ and $\bar{c}$ each represent $-1$ degrees of freedom.  $c$ exists to cancel out the pure gauge modes $A = \nabla \alpha$, while $\bar{c}$ exists to cancel out the Lorenz-gauge violating modes with $\nabla_a A^a \ne 0$.  

The partition function can then be expressed as a functional determinant
\begin{equation} \label{maxwelldeterminant}
\ln Z = -\frac{1}{2} \ln \det \Delta_1 + \ln \det{\Delta_0}.
\end{equation}
where
\begin{equation}
\Delta_0 = -\nabla^2, \qquad 
\Delta_1 = -g_{ab} \nabla^2 + R_{ab}.
\end{equation}
The ghosts are minimally coupled and so do not contribute a contact term.

The manipulations leading to Eq.~\eqref{Scurv} can be repeated for any theory in which the action depends on the Riemann tensor.
On the conical manifold, the singular part of the Riemann tensor is given by \cite{Fursaev1995}
\begin{equation}
R^\mathrm{tip}_{abcd}(x) = (2 \pi - \beta) \epsilon_{ab} \epsilon_{cd} \delta_\Sigma (x).
\end{equation}
The calculation proceeds much as in the case of the nonminimally coupled scalar, with the result that
\begin{equation}
S_\text{tip} = -2 \pi \avg{\int_\Sigma d^{D-2}x \sqrt{h} \frac{\partial L}{\partial R_{abcd}} \epsilon_{ab} \epsilon_{cd} }.
\end{equation}

Calculating the Wald entropy \eqref{wald} by differentiating the action \eqref{action} with respect to the curvature, we obtain \cite{privateKabat}:
\begin{equation}\label{A^2}
S_\mathrm{Wald} = -\pi \int_\Sigma d^{(D-2)}x\,\sqrt{h}\,A_a A_b g_\perp^{ab},
\end{equation}
where $g_\perp^{ab}$ is the inverse metric projected onto the directions perpendicular to the horizon.
The expectation value of Eq.~\eqref{A^2} is the same as the contribution of two nonminimally coupled scalars with $\xi = 1/2$ \cite{Larsen1995}. This gives a contribution $c_1 = -1$, in exact agreement with the contact term.
Thus at first sight it appears that the antiscreening of Newton's constant can be explained physically through a divergence in the Wald entropy.

But this interpretation is problematic, because the $A^2$ term is not gauge-invariant. This lack of gauge invariance is just a consequence of the gauge-fixing of the action \eqref{action}.
However an avatar of the original gauge invariance remains in the form of the fermionic BRST symmetry $\mathbf{s}$ relating the ghosts to the unphysical vector modes:
\begin{align}\label{BRST}
\mathbf{s} A_a = \nabla_a c, \qquad
\mathbf{s} \bar{c} = \nabla_a A^a,\qquad
\mathbf{s} c = 0.
\end{align}
BRST symmetry guarantees that the expectation values of BRST-invariant operators are independent of the choice of gauge.
The operator appearing in Eq.~\eqref{A^2} is not BRST-invariant, but instead transforms as
\begin{equation}
\mathbf{s} (A_a A_b g_\perp^{ab}) = 2 A_a \, \nabla_b c \, g_\perp^{ab}.
\end{equation}
This may help explain the results of Ref.~\cite{Iellici1996}, where it was found that the contact term depends on the parameter $\xi$ of the $R_\xi$ gauge using $\zeta$-function regularization in $D=4$ (though not in $D=2$).

The $A^2$ term \eqref{A^2} is actually a JKM ambiguity \cite{Jacobson1993} in the definition of the Wald entropy, since it can be removed by adding a total derivative to the Lagrangian; the Wald entropy \eqref{wald} of the original Maxwell Lagrangian $\frac{1}{4}F_{ab} F^{ab}$ vanishes.  
Classically, ambiguity terms such as $A_a A_b g^{ab}_\perp$ vanish on the Killing horizon for stationary field configurations, but in the quantum theory they can have nonzero expectation values.

Additionally, since Maxwell fields (coupled to general relativity) satisfy the null energy condition, there is a classical second law in which the horizon entropy is given by the Bekenstein-Hawking area term alone.  The addition of Eq.~\eqref{A^2} to the entropy seems likely to spoil this result.
This is in contrast with the nonminimally coupled scalar field, for which the inclusion of the Wald entropy term $- 2\pi \xi \phi^2$ is necessary for the classical second law \cite{Ford2000}.

\section{Derivation of the Kabat contact term}\label{kabat}

We now summarize the calculation that led to the puzzling contact term in Eq.~\eqref{c1}.

Following the same method as in section \ref{renWald}, we express the partition function in terms of the heat kernels of the vector and scalar Laplacians.
Let $\phi_n$ be a complete set of modes for $\Delta_0$ (with $\xi = 0$):
\begin{equation}
-\nabla^2 \phi_n = \lambda_n \phi_n.
\end{equation}
The scalar heat kernel is given by
\begin{equation}
K_S(s,x,y) = \sum_{n} e^{-s \lambda_n} \phi_n(x) \phi_n(y).
\end{equation}
Although we have written the heat kernel in the case of a discrete spectrum, the results generalize naturally to the case of continuous spectrum.

To compute the heat kernel of the vector Laplacian, we construct a complete set of eigenfunctions of the operator $\Delta_1$:
\begin{equation}
(-g_{ab} \nabla^2 + R_{ab}) A^b = \lambda_n A^a,
\end{equation} 
and define the vector heat kernel
\begin{equation}
K_{V}(s,x,y)_{ab} = \sum_n e^{-s \lambda_n} A_{na}(x) A_{nb}(y).
\end{equation}
In two dimensions, the vector modes can be written in terms of the scalar eigenfunctions as 
\begin{equation}
\frac{1}{\sqrt{\lambda_n}} \nabla_a \phi_n, \qquad \frac{1}{\sqrt{\lambda_n}} \epsilon_{ab} \nabla^b \phi_n.
\end{equation}
The vector heat kernel at coincident points and with the vector indices contracted can be expressed in terms of the scalar heat kernel as
\begin{align}
K_V(s,x,x)_a^a &= \sum_n \frac{e^{-s \lambda_n}}{\lambda_n} [ 2 \nabla_a \phi_n \nabla^a \phi_n] \\
&= \sum_n  \frac{e^{-s \lambda_n}}{\lambda_n} [- 2 \phi_n \nabla^2 \phi_n + 2 \nabla_a (\phi_n \nabla^a \phi_n)] \\
&= \sum_n e^{-s \lambda_n} [2 \phi_n^2 + \frac{1}{\lambda_n} \nabla^2 (\phi_n^2)] \\
&= 2 K_S(s,x,x) + \int_s^\infty ds' \nabla^2 K_S(s',x,x). \label{vectorscalar}
\end{align}
In the heat kernel regularization, the partition function \eqref{maxwelldeterminant} is given by
\begin{equation}
\ln Z = \frac{1}{2} \int_{\epsilon^2}^\infty ds \frac{e^{-m^2 s}}{s} K_M(s)
\end{equation}
where the mass $m$ is an infrared regulator. 
Here we have defined $K_M(s)$ as the trace of the ``Maxwell heat kernel'':
\begin{align} \label{KMdef}
K_M(s) &=  \int d^2x \sqrt{g} \left( K_V(s,x,x) - 2 K_S(s,x,x) \right) \\
&= \int d^2x \sqrt{g} \int_s^\infty ds' \nabla^2 K_S(s',x,x) \label{KMint}
\end{align}
using \eqref{vectorscalar} in the last line.
It seems tempting to move the integration with respect to $s$ past the Laplacian, turning this expression into the integral of a total derivative.
However this would not be valid as $K_S$ behaves as $1/s$ for large $s$, so the $s$ integral would be ill-defined.

To evaluate $K_M$ on a cone of angle $\beta$, Kabat exploits rotation and scale symmetry to write $K_S(s',x,x) = f(r^2/s')/s'$, so that Eq.~\eqref{KMint} becomes
\begin{align}
K_M(s) &= \beta \int r dr \int_s^\infty ds' \frac{1}{r} \partial_r r \partial_r s' f(r^2/s').
\end{align}
By dimensional analysis, $r \partial_r K = - 2 \partial_s (s K)$. Both integrals can then be carried out, yielding
\begin{equation} \label{orderoflimits}
K_M(s) =  -2 \beta  f(r^2/s') \big|_{s'=s}^{s'=\infty} \big|_{r=0}^{r=\infty}.
\end{equation}
When $r^2 \gg s$, the heat kernel on a cone takes the same form as on the plane, $f(r^2/s) \approx \frac{1}{4 \pi}$.
When $r^2 \ll s$, the heat kernel is very sensitive to the conical singularity, and $f(r^2/s) \approx \frac{1}{2 \beta}$.
In Eq.~\eqref{orderoflimits}, there are two contributions from $r = 0$ that each give $\frac{1}{2 \beta}$, and a contribution from $r = \infty$, $s'=s$ that yields $\frac{1}{4\pi}$.
But there is also a contribution from $s = \infty$, $r = \infty$ that depends on the order in which the limits are taken.
If we take the limit $s \to \infty$ first, we find the same result as Kabat:
\begin{equation} \label{KM}
K_M(s) = -\frac{1}{2 \pi} ( 2 \pi - \beta) = - \frac{1}{4 \pi} \int \sqrt{g} R.
\end{equation}
If instead we take the $r \to \infty$ limit first, the Maxwell heat kernel vanishes identically, and we obtain no contact term.

The partition function associated to \eqref{KM} is given by
\begin{equation}
\ln Z = -\frac{1}{4 \pi} (2 \pi - \beta) \int_{\epsilon^2}^\infty ds \frac{e^{-m^2 s}}{s},
\end{equation}
from which we easily find the entropy using \eqref{SfromZ}:
\begin{equation}
S_\text{cone} = -2 \pi \int_{\epsilon^2}^\infty ds \frac{e^{-m^2 s}}{4 \pi s}.
\end{equation}
This corresponds to a term $c_1 = -1$ in the conical entropy \eqref{Sc1}.

We note that the Maxwell heat kernel on the cone \eqref{KM} is independent of $s$.
Thus it enters the heat kernel in the same way as $(\beta/2 \pi - 1)$ zero modes would, although this coefficient is not in general an integer.
This suggests that the calculation may depend on the way that zero modes are handled.  Indeed, the dependence on the order of limits $s \to \infty$ and $r \to \infty$ shows that the calculation is sensitive to the far infrared.
Taking the limit $s \to \infty$ first corresponds to allowing paths a sufficient amount of Schwinger proper time to probe the boundary at large $r$.  

In section \ref{zeromodes} we will repeat the contact term calculation on a smooth compact space without boundary or singularities.  
We will see that the contact term does indeed arise from zero modes.

\section{2D Maxwell theory on a compact spacetime}\label{zeromodes}

Because Kabat derived the contact term on a manifold with a conical singularity and a boundary at infinity, one might wonder whether the result comes from the improper treatment of these boundaries. 
To show that this is not the case, in this section we will re-derive the Kabat contact term for smooth compact orientable two-dimensional Euclidean manifolds.  However the interpretation is different: the contact term in the entropy arises from zero modes of ghosts, explaining its negative sign.

When calculating the entropy on smooth manifolds, we will replace the conical singularity with a smooth cap, smearing out the curvature over some finite radius $r_0$ \cite{Nelson1994,Fursaev1995,Frolov1995}.  Because of approximate translation symmetry near the horizon, to first order in the angle deficit $2\pi - \beta$ and in the limit that $r_0 \to 0$, the heat kernel does not depend on the details of the smoothing.  Formally therefore, the replacement of the conical singularity with the smoothed tip should have no consequences, and indeed this is what we will find.

To compute the effective action, we use the trace of the Maxwell heat kernel \eqref{KMdef}
\begin{equation}
K_M(s) = \mathrm{tr} ( e^{-s \Delta_1} ) - 2 \, \mathrm{tr} ( e^{-s \Delta_0} ).
\end{equation}
Both operators $\Delta_0$ and $\Delta_1$ are cases of the Hodge Laplacian acting on $p$-forms:
\begin{equation}
\Delta_p = d \delta + \delta d
\end{equation}
where $d$ is the exterior derivative and $\delta$ is the codifferential.

By the Hodge decomposition, any $1-$form $A$ can be expressed as
\begin{equation} \label{hodge}
A = d \phi + \delta \psi + B
\end{equation}
where $\phi$ is a $0$-form (scalar), $\psi$ is a 2-form and $B$ is a harmonic 1-form i.e. $\Delta_1 B = 0$.
By Eq.~\eqref{hodge}, the spectrum of $\Delta_1$ is the union of the spectrum of $\Delta_0$ and $\Delta_2$ up to zero modes.
Moreover, by Hodge duality, the spectra of $\Delta_0$ and $\Delta_2$ are equivalent on orientable manifolds.
In terms of the heat kernels, this implies that 
\begin{equation} \label{Vequals2S}
K_V(s) = 2 K_S(s) + b_1 - b_0 - b_2
\end{equation}
where $b_p = \dim \ker \Delta_p$ is $p^\text{th}$ Betti number, which counts the number of $p-$form zero modes.
On a connected orientable manifold, $b_0 = b_2 = 1$ and $b_1$ is twice the genus, so we have
\begin{equation} \label{KV}
K_V(s) = 2 K_S(s) - \chi
\end{equation}
where $\chi = b_0 - b_1 + b_2$ is the Euler characteristic.

Subtracting the two scalar ghosts from the vector heat kernel \eqref{KV}, we find the Maxwell heat kernel
\begin{equation} \label{gaussbonnet}
K_M(s) = -\chi = - \frac{1}{4 \pi} \int d^2x \sqrt{g} R.
\end{equation}
where we have used the Gauss-Bonnet theorem.
Now note the similarity between this result and Kabat's result for the cone: the right-hand side of Eq.~\eqref{gaussbonnet} and Eq.~\eqref{KM} are the same.

To find $\ln Z$ in terms of the heat kernel, we again introduce an ultraviolet cutoff length $\epsilon$ and an infrared regulating mass $m$,
\begin{eqnarray}
\ln Z &=& \frac{1}{2} \int_{\epsilon^2}^\infty ds \frac{e^{-m^2 s}}{s} K_M(s)\\
&=& -\int \sqrt{g} 
\left( \frac{1}{2} \int_{\epsilon^2}^\infty ds \frac{e^{-m^2 s}}{4 \pi s} R \right).
\end{eqnarray}
By comparison with equation \eqref{Leff}, this corresponds to $c_1 = -1$ in the effective action.

We have therefore confirmed the presence of the contact term for compact manifolds.
But more importantly, we have elucidated the origin of the contact term: it arises from the difference in the number of degrees of freedom in the vector zero modes as compared to the ghost zero modes.\footnote{One may wonder how these zero modes could possibly give rise to a logarithmic divergence in $1/G$, considering that a finite number of modes cannot give rise to an ultraviolet divergence.  The explanation is that when taking the determinant of a dimensionful operator $\Delta$, one must insert a dimensionful parameter $\mu$ so that the partition function $Z = \mathrm{det}(\mu^{-2} \Delta)$ is dimensionless.  Although conceptually $\mu$ has no necessary relationship to an ultraviolet cutoff $\Lambda$ on short-distance modes, since both $\mu$ and $\Lambda$ are dimensionful parameters needed to make the path integral well defined, one may as well identify $\Lambda = \mu$.  In any case both parameters must be varied in order to perform an RG flow.  The distinction between these two conceptually distinct reasons for renormalization is obscured by the heat kernel regulator $\epsilon$, which does not distinguish between them.}

As an example, let us consider the effective action on de Sitter space $dS_2$, for which the Euclidean geometry is the sphere $S_2$.  All modes cancel except for the two zero modes of the ghosts $c$ and $\cbar$, since there are no vector zero modes on the sphere.
These modes are ghosts and contribute negatively to the entropy $S = (1 - \beta \partial_\beta) \ln Z$.  The $\beta$ variation of the sphere (which corresponds to deforming it into a ``football'' with two smoothed out conical caps) vanishes, because $Z$ depends only on the topology, not $\beta$.  This leaves only the $\ln Z$ term, which is negative.

Thus we rederive the contact term, but now the interpretation is that it comes from negative entanglement entropy due to ghosts.  But this should immediately arouse suspicion! 
The sole purpose of ghosts is to cancel out unphysical vector modes, and the ghost zero modes are extra fields which are not associated with any such vector modes.

A similar calculation was carried out in Ref.~\cite{Barvinsky1995} in which all vector and ghost zero modes were omitted, leading to a trivial partition function $Z=1$.
This prescription removes the contact term; however it neglects nonperturbative contributions to $Z$ that will be considered in section \ref{physical2D}.

\subsection{Problems with the na\"ive Maxwell heat kernel}\label{problems}

Let us go back to the original justification for the ghosts.  In the Faddeev-Popov trick, one takes a path integral of the form
\begin{equation}
\int \mathcal{D} A\,e^{-S(A)},
\end{equation}
and inserts the ``identity''
\begin{equation} \label{fptrick}
\int \mathcal{D} \alpha \, \delta(G(A^\alpha)) \det \left[ \frac{\delta G(A^\alpha)} {\delta \alpha} \right] = 1.
\end{equation}
Here $G(A) = \nabla_a A^a$ is the Lorenz gauge-fixing condition, and $A^\alpha = A + \nabla \alpha$.
This assumes that for every $A$ there is exactly one $\alpha$ such that $G(A^\alpha) = \nabla_a A^a + \nabla^2 \alpha = 0$.
However if $\alpha$ satisfies this condition, then so does $\alpha + c$ where $c$ is a spacetime constant.
This means that we should integrate over equivalence classes of functions $\alpha$ under the relation $\alpha \sim \alpha + c$, in other words the determinant in Eq.~\ref{fptrick} should not include zero modes; hence the ghost zero modes are spurious.

On manifolds with handles, the vector zero modes must also be treated with great care.  If the gauge group of the Maxwell theory is $\mathbb{R}$, there will be infrared divergences coming from these winding modes.  For a U(1) gauge field, this infrared divergence is replaced with an integral over the moduli space of flat connections, which has finite volume.  These zero modes must be exluded from the one-loop determinant and handled separately.  It is also necessary to sum over nontrivial U(1) bundles.

Finally, there is an additional problem that the BRST state space is not the same as the physical Hilbert space of the canonical Maxwell theory, but contains extra degrees of freedom with negative norm states.
This problem arises on any spatially compact manifold, but for specificity consider a static Lorentzian manifold (of any dimension) taking the form $\Sigma \times \mathbb{R}_\mathrm{time}$.  Let $q$ be the determinant of the spatial metric, and $t$ be the time coordinate on $R$.  On any $t = \mathrm{const.}$ time slice, the following pair of canonically-conjugate spatially constant ghost modes
\begin{equation}
c_0 = \int_\Sigma d^{D-1}x \sqrt{q}\,c, \quad \dot{\bar{c}}_0 = \int_\Sigma d^{D-1}x \sqrt{q}\,\frac{d\bar{c}}{dt}
\end{equation}
are BRST-trivial, i.e. they are not paired by BRST symmetry with any other modes. 
In the canonical BRST formalism, the physical Hilbert space is defined as the cohomology of $Q$, the generator of the BRST symmetry $\mathrm{s}$ \eqref{BRST}.
In other words, the Hilbert space is given by restricting to states in the kernel of $Q$, and modding out by states in the image of $Q$.
This means that the BRST-trivial ghost modes remain in the ``physical'' state space despite the fact that they include negative norm states, and do not correspond to any modes in the canonical Maxwell theory.  These spurious degrees of freedom are similar to the extra ghosts that arise when BRST-quantizing the zero mode of a string, and which are normally cured by imposing Siegel gauge \cite{Siegel1988}.  This problem arises even in the 0+1 dimensional gauge-fixed Maxwell theory, where there are are two Faddeev-Popov ghosts, yet only one component of the vector field.

These problems cast doubt on the validity of the ``vector minus two scalars'' calculations of the contact term.  Rather than try to resolve these issues here, we will instead quantize the two-dimensional theory using the reduced phase space of gauge-invariant canonical degrees of freedom.  We will see that the contact term is absent.

\section{Reduced phase space quantization}\label{physical2D}

Two dimensional Maxwell theory has no local degrees of freedom, but there are still global degrees of freedom. In fact, the system is exactly solvable without the introduction of gauge fixing or ghosts (for a review see Ref.~\cite{Cordes1994}).

On a 2-dimensional orientable Euclidean manifold the Maxwell action is
\begin{equation} \label{2daction}
I = \int d^2x \sqrt{g} \frac{1}{2} F^2,
\end{equation}
where we define $F = \frac{1}{2 \sqrt{g}} F_{ab}\epsilon^{ab}$.
In order to perform a canonical analysis, we will start by assuming that the manifold can be foliated by circles (i.e. is a sphere or torus); this assumption will be lifted at the end of this section.

At a fixed time, the configuration degrees of freedom are the gauge equivalence classes of a $\mathrm{U}(1)$ connection $A_a$ on the circle.
These equivalence classes are parameterized by a single degree of freedom, the Wilson loop around the circle,
\begin{equation}
A = \oint A_a dx^a.
\end{equation}
Note that the action \eqref{2daction} depends on the metric only via the volume form.
By choosing a coordinate $x \in [0,1]$ parameterizing the circle, and a coordinate $t$ that measures the elapsed spacetime volume, the volume element becomes $\sqrt{g} d^2x = dt dx$.
Then we can reduce the phase space by imposing Coulomb gauge, in which $A_x$ is constant, and $A_t = 0$.
Maxwell theory in two dimensions then reduces to the free particle with Hamiltonian
\begin{equation}
H = \frac{1}{2}E^2
\end{equation}
where the electric field $E$ is canonically conjugate to $A$:\footnote{
Although on-shell $E = F$, off-shell it is important to distinguish between the momentum $E$ and the velocity $F$.  The former is conserved and the latter fluctuates.} 
\begin{equation} \label{2dpoisson}
\{ A,E \} = 1.
\end{equation}
All the relevant information about the manifold is encoded in the total volume $V$, and the boundary conditions imposed on $A$ at $t = 0$ and $t = V$.

To quantize the theory, we simply replace the Poisson brackets by commutators, giving a free particle.
For a theory with gauge group $\mathbb{R}$, $A$ can take any real value, but for a U(1) gauge theory $A$ is periodic:
\begin{equation}
\quad A \sim  A + \frac{2\pi}{q},
\end{equation}
where $q$ the minimal charge, so the free particle lives on a circle.
This implies that the electric field is quantized as
\begin{equation}
E \in q \mathbb{Z}.
\end{equation}

To compute the partition function, we first need to specify the topology of the Euclidean manifold, which determines the boundary conditions for $A$.
We first consider the torus, for which the appropriate boundary conditions are the periodic ones:
\begin{equation}
A(0) = A(V).
\end{equation}
The partition function is
\begin{equation}\label{Epath}
Z = \mathrm{tr}\,e^{-V H}
= \sum_{E \in q \mathbb{Z}} e^{-\frac{1}{2} V E^2}.
\end{equation}

We can also compute $Z$ by the Euclidean path integral.
Because the action is quadratic, we can factor the partition function into a sum over classical paths times a contribution from fluctuations about the classical paths,
\begin{equation} 
Z = \sum_{n \in \mathbb{Z}} e^{-S[A_n]} \times Z_\mathrm{fluctuations}
\end{equation}
where $A_n$ is the classical path that wraps around the circle with winding number $n$: 
\begin{equation}
A(t) = \frac{2 \pi n}{q V} t, \qquad F = \frac{2 \pi n}{q V}
\end{equation}
which is the familiar quantization of magnetic flux.
The fluctuations can be calculated from the Euclidean free particle propagator on the plane,
\begin{equation}
U(\Delta x, \Delta \tau) = \sqrt\frac{1}{2 \pi \Delta \tau} e^{-(\Delta x)^2/2t},
\end{equation}
yielding
\begin{equation} \label{Zfluctuations}
Z_\textrm{fluctuations} = \int_0^{2 \pi/q} dA \, U(0, V) = \sqrt\frac{2 \pi}{q^2 V}.
\end{equation}
Combining this result with the classical action, the partition function is
\begin{equation} \label{Zeuc}
Z = \sqrt\frac{2 \pi}{q^2 V} \sum_{F \in (2 \pi/q V) \mathbb{Z}} e^{-\tfrac{1}{2} V F^2}.
\end{equation}
While the formulae for the partition function Eq.~\eqref{Epath} and Eq.~\eqref{Zeuc} have a similar form, the quantization of $E$ (electric quantization) and of $F$ (magnetic quantization) are completely different in nature.
The electric quantization condition is quantum kinematical effect arising from the finite radius of the circle, whereas the magnetic quantization condition is a classical topological result that makes use of the equation of motion.
Nevertheless, Eq.~\eqref{Epath} and Eq.~\eqref{Zeuc} can be shown to be equal by the Poisson summation formula.

When the spacetime manifold is a sphere, the circle shrinks to a point at $t = 0$ and $t = V$, leading to the boundary conditions
\begin{equation}
A(0) = A(V) = 0.
\end{equation}
The partition function on the sphere is then given by 
\begin{equation}\label{bound}
Z = \bra{\psi} e^{-V H} \ket{\psi}
\end{equation}
where $\ket{\psi}$ is the (unnormalizable) wavefunction given in the $E$ basis by $\psi(E) = 1$ and in the $A$ basis by $\psi(A) = \sqrt{2\pi/q} \delta(A)$.\footnote{Although this normalization is the most natural, if one were to choose a different normalization of $\ket{\psi}$, this would be equivalent to a finite shift of $1/G$.}
The result is identical to Eq.~\eqref{Epath}, showing that the partition function of 2D Maxwell theory does not distinguish between a sphere and a torus of the same volume.

In fact, we can generalize this result to Euclidean manifolds of arbitrary genus by sewing together manifolds with boundary.  Using the same boundary condition \eqref{bound} as the sphere, one can show that manifold of volume $V$ with the topology of a disk produces the state $\psi(E) = e^{-\frac{1}{2} V E^2}$.  These disks can be sewn together using manifolds with three spatial boundaries (``pairs of pants'').  If we consider a pair of pants in the limit in which the volume vanishes, it can be viewed as a wavefunction of the electric fields on each of its three boundaries, given by $\psi(E_1, E_2, E_3) = \delta(E_1, E_2) \delta(E_2, E_3)$, where the normalization factor is fixed by the requirement that one recover the partition function of the sphere when sewing the pants to three disks.
By sewing together an arbitrary number of pants and disks, we find that the partition function for an arbitrary two-dimensional closed Euclidean manifold without boundary depends only on the volume,
\begin{equation} \label{Zcan}
Z = \sum_{E \in q \mathbb{Z}} e^{-\frac{1}{2} V E^2}.
\end{equation}

The conical entropy is easily calculated from Eq.~\eqref{Zcan}. 
Since the volume of the Euclidean manifold is linear in the deficit angle $\beta$, the formula \eqref{SfromZ} for the entropy yields
\begin{equation} \label{2DGibbs}
S = (1 - V \partial_V) \ln Z = - \sum_{E \in q \mathbb{Z}} p(E) \ln p(E)
\end{equation}
where $p(E)$ is the probability of measuring a given value of $E$ locally, $p(E) = Z^{-1} e^{-\frac{1}{2} V E^2}$.
This entropy is manifestly positive, and has an obvious statistical interpretation: the only local observable is $E$, and this is constant over space.  Therefore observers on different sides of the horizon measuring $E$ will find perfect correlation of their measurement results; the degree to which their states are entangled is given by the entropy in Eq.~\eqref{2DGibbs}.
We conclude that in two dimensions, the conical entropy of a gauge field coincides with its entanglement entropy.
Note that this entropy vanishes in the large volume limit $q^2 V \to \infty$.

The results of this section can be generalized immediately to $(D - 1)$-form electromagnetism in $D$ dimensions.  Since the action depends only on the total spacetime volume, the dimension is irrelevant.

Although $1/G$ is not renormalized in the reduced phase space space
calculation, one might worry that this may depend on the quantization
scheme used.  
In a $D = 2$ theory with scale-invariance, any divergence in $1/G$ would be logarithmic, with a scheme-independent coefficient which appears in the trace anomaly \cite{DiFranchesco1997}.  
However, $D = 2$ electromagnetism is not scale-invariant since the minimum charge $q$ is dimensionful.
However, we note that the trace of the stress-energy tensor is scheme-independent.
In the reduced phase space calculation, this is given by
\begin{equation}
\avg{T} = - \avg{E^2}.
\end{equation}
This is equal to the classical result, and does not include any term proportional to the Ricci curvature $R$, as expected in a theory in which $1/G$ is not renormalized.

\subsection{Topological Susceptibility}\label{top}

In Ref.~\cite{Zhitnitsky2011}, it was proposed that the contact term is related to the topological susceptibility $\chi_\mathrm{t}$, which measures the response of $F$ to the introduction of a source term 
$i (\frac{q}{2 \pi}) \theta \int \sqrt{g} F$,
\begin{equation}
\chi_\mathrm{t} 
= -\left . \frac{1}{V}\frac{\partial^2}{\partial \theta^2} \ln Z \right|_{\theta = 0}
= Z^{-1} \sqrt{\frac{2 \pi}{q^2 V}} \sum_{F \in \frac{2 \pi}{q V} \mathbb{Z}} \left( \frac{q}{2 \pi} \right)^2 V F^2 e^{-\frac{1}{2} V F^2}
 = \left( \frac{q}{2 \pi} \right)^2 V \avg{F^2}.
\end{equation}
The topological susceptibility has properties reminiscent of Kabat's contact term: in particular the contribution from the electromagnetic field has a sign opposite to all possible matter terms (which contribute negatively).
In Ref.~\cite{Zhitnitsky2011} it was conjectured that the ``wrong sign'' term in the topological susceptibility is responsible for the negative contact term in the entropy.
We will now show that, although the entropy remains manifestly positive, the topological susceptibility does contribute to the entropy with a negative sign.

To see how the susceptibility appears in the entropy, we can compute the entropy from the partition function \eqref{Zeuc}:
\begin{equation}
S = (1 - V \partial_V) \ln Z = \ln Z + \frac{1}{2} - \frac{1}{2} \left( \frac{2 \pi}{q} \right)^2 \chi_\mathrm{t}.
\end{equation}
The first term is proportional to the free energy.
The $1/2$ comes from the fluctuation term \eqref{Zfluctuations}. The last term comes from differentiating the sum over nontrivial bundles, and is proportional to the topological susceptibility. 
It appears that the $\chi_\mathrm{t}$ term could make the entropy negative, but this is not the case. 
At small $V$, the $\ln Z$ term is positive and dominates the entropy.
As $V$ increases, $\chi_\mathrm{t}$ increases to $(q/2\pi)^2$ in the large volume limit, and its negative contribution to the entropy exactly cancels with the $1/2$ contribution from the fluctuations.

\subsection{Yang-Mills}

The partition function of non-abelian gauge theory is also known exactly in 2D and is given by the generalization of Eq.~\eqref{Zcan},
\begin{equation}
Z = \sum_{R} (\dim R)^\chi e^{-\frac{1}{2} q^2 V C_2(R)}
\end{equation}
where the sum extends over all irreducible unitary representations $R$ of the gauge group, and $C_2$ is the quadratic Casimir.
In the U(1) theory, the representations are labelled by integers, with $\dim(R_n) = 1$ and $C_2(R_n) = n^2$.
The dependence on the Euler characteristic $\chi$ is reminiscent of the contact term, but the contribution of this term to the entropy is finite and positive:
\begin{equation}
S = -\sum_R p(R) \ln p(R) + \sum_R p(R) \ln \dim R
\end{equation}
where $p(R)$ is the probability distribution over representations, $p(R) \propto e^{-\frac{1}{2} q^2 V C_2(R)}$.
We can see that the entropy is positive, and is equal to the entanglement entropy derived in Ref.~\cite{Donnelly2011}.

\section{Discussion}\label{dis}

We have shown in section \ref{physical2D} that in two dimensions, when the physical Wilson loop is regulated by introducing a finite minimum charge and is normalized correctly, the partition function depends only on the Euclidean volume, not the curvature. 
The contribution of the Maxwell field to the effective action is finite, and vanishes as the Euclidean volume goes to infinity.
In two dimensions, Maxwell fields do not renormalize $1/G$.
The contact term does not appear when the theory is quantized using the true physical degrees of freedom.  Hence there is no need to include the gauge-dependent term in the Wald entropy discussed in section \ref{aint}.  Thus, the JKM ambiguity in the Wald entropy is resolved in a gauge-invariant way: the contribution to the Wald entropy from a Maxwell field is zero.

This result is in disagreement with the partition function computed from the Maxwell heat kernel $K_M = K_V - 2K_S$, which we have calculated for compact manifolds in section \ref{zeromodes}.  Although we confirm the existence of the contact term in this model, the model is unphysical because it includes contributions from spurious ghosts identified in section \ref{problems}. 
The path integral contains a contribution from ghost zero modes, and the canonical phase space contains a pair of spatially constant BRST-invariant ghost modes.
Additionally, the infrared divergence of the vector zero modes was treated in an unphysical way by introducing a small mass.  This is physically incorrect since gauge fields cannot be given masses without introducing an extra degree of freedom.  The physically correct infrared regulator is invariance under large U(1) gauge transformations, and this gives a different result for the partition function.

Since a noncompact manifold can be viewed as the limit of an infinitely large compact manifold, the absence of the contact term ought to manifest somehow in this limit as well.  Since a noncompact manifold has a continuous spectrum, it is harder to see the effects of the zero mode prescription.  However, in section \ref{kabat} it was observed that the derivation of the contact term for the cone is sensitive in the infrared to an order of limits: if one takes $r \to \infty$ before taking $s \to \infty$, the contact term does not appear.  Thus the calculation on the cone is also sensitive to the prescription for dealing with the infrared aspects of the theory.

Although these conclusions are confined to the case of $D = 2$, the absence of the contact term in $D = 2$ suggests that the $D > 2$ calculations should also be revisited.  
Since in Kabat's derivation, the contact term in the higher-dimensional heat kernel just comes from the product of the contact term in the two-dimensional Maxwell heat kernel times the $D - 2$-dimensional scalar heat kernel,
one might think that the contact term will also be absent in higher dimensions. However, in higher dimensions the contact term no longer arises solely because of zero modes, so the analysis will be different.  Since $1/G$ is power-law divergent in $D > 2$, the results may also depend on ones choice of renormalization scheme, as well as the choice of gauge \cite{Iellici1996}.

A similar negative contact term appears in the case of gravitons \cite{Fursaev1996,Solodukhin2011}.  Hence pure gravity seems to antiscreen itself, suggesting the possibility of a nontrivial UV fixed point at positive $G$ \cite{Niedermaier2006}.  However, since the Maxwell contact term is not actually present (at least in $D = 2$), these calculations should be carefully revisited. 

\subsection*{Acknowledgements}
\small
We are grateful for discussion with Ted Jacobson, Dan Kabat, Sergey Solodukhin, Joe Polchinski, Don Marolf, Bernard de Wit, Amanda Peet, and Ashoke Sen.
AW is supported by NSF grants PHY-0601800, PHY-0903572, the Maryland Center for Fundamental Physics, and the Simons Foundation.
WD is supported by the Foundational Questions Institute (Grant No.
RFP20816), NSERC, and the Perimeter Institute.
We are both grateful for the support of the Kavli Institute for Theoretical Physics while this work was being completed.
\normalsize

\bibliographystyle{utphys}
\bibliography{max2d}

\end{document}